# On the Capacity of p2p Multipoint Video Conference

Zhao Yong-xiang
School of Electronics and Information
Beijing Jiaotong University
Beijing 100044, China
e-mail: yxzhao@bjtu.edu.cn,

Chen chang-jia
School of Electronics and Information
Beijing Jiaotong University
Beijing 100044, China
e-mail:changjiachen@sina.com

*Abstract*—In this paper, The structure of video conference is formulated and the peer-assisted distribution scheme is constructed to achieve optimal video delivery rate in each sub-conference. The capacity of conference is proposed to referee the video rate that can be supported in every possible scenario. We have proved that, in case of one user watching only one video, 5/6 is a lower bound of the capacity which is much larger than 1/2, the achievable rate of chained approach in [2]. Almost all proofs in this paper are constructive. They can be applied into real implementation directly with a few modifications.

*Keywords-video conference*

## I. INTRODUCTION

With the proliferation of the Internet, video conferences are deployed widely. For instance, the latest version of SKYPE [1] has enabled function of multipoint (MP) video conference which requires large bandwidth. One approach to reduce the bandwidth demand of MP video conference is to use a device called multipoint control unit(MCU) to collect all participants' video and disseminate videos to all participates of conference. However, since all the traffic has to pass through the access link of MCU, operation of MCU will lead to huge cost and management complexities. Another way is to use peer-to-peer (P2P) approach for MP conference [2]. In this approach, a participant may have to relay the received video signal to another participant who also wants the same signal and a chain like path consisted by participant who need a same video signal is formed to relay the signal. [3] extends this idea to multilayer scheme. [6] uses End System Multicast protocol to transfer the data in video conference and support multiple video source.

The video quality or equivalent the achievable video rate can be supported by a p2p MP video conference is important in designing a MP video conference system and also in evaluating an implementation of MP video conference system. However, this problem has never been exploited based on our knowledge. The achievable rate and the capacity in a MP video conference will be addressed in this paper.

Incented by works in [4, 5], we borrow the optimal mechanism of file distribution in [4] into MP video conference to solve the bandwidth demand requirement. In section 2, the structure of videoconference is studied, and Peer-Help Module (PHM), which is a scheme to assign users uplink bandwidth among different video source, is also established. In section 3, Peer-assisted video distribution based on the PHM is explored to find the achievable video rate in MP conference, which leads to following finding: the achievable rate of a video source varies with the number of users who are viewing the video. However, adaption of the video rate according to the number of its viewers is not feasible in implementation since the video rate cannot be quickly changed in most existing video compression algorithms; Thus, the concept of capacity of conference is proposed to calculate the video rate that can be supported in every possible scenario. In other words, if each video source generates it's video at rate given by conference capacity, no matter how dynamically the users change their watching relationship, suitable PHM module can be quickly identified to guarantee the delivery of videos from senders to their watchers. In this paper, we have also proved that, in case of one user watching only one video, 5/6 is a lower bound of the capacity which is much larger than 1/2 that is the achievable rate implied in the chained approach in [2].

Both PHM and the distribution scheme in the PHM are constructive and can also be applied directly in the practical implementation with a few modifications.

## II. THE STRUCTURE OF A CONFERENCE AND THE PEER-HELP MODULE

In this paper, Symbol $C(N, k)$ represents a conference in which has $N$ users and each user is watching exactly $k$ videos generated by other users at any given instant. We also assume one user can generate a video at most. For example, [6] claims that in awareness-driven 3D videoconferencing application, a user on average watches one or two videos of other users' since the field of view does not allow for three users to be simultaneously in view. In general, a conference is a dynamic process, which means one user could switch among videos generated by other users. Thus the watching relationships between the video source and its viewer may change from time to time. Snapshot of watching relationships among conference users is named as the scenario at a given time instance. It is worthwhile to notice that, the number of different relationship patterns in scenarios is limited. Specifically, a scenario $S$ can be specified by a sequence of $\{G_i: 1 \leq i \leq N\}$, where $G_i$ is the subset of users who are watching the video generated by $i^{th}$ user. For simplicity, we will use video $i$ to name the video generated by $i^{th}$ user. Since $G_i$ is the group of users watching video $i$, we will also name $G_i$ as the **sub-conference** $i$ in the scenario. In each scenario of $C(N, k)$ conference, every user is exactly in $k$ sub-conferences at same time. In this paper, the number of elements in a given set $G$ is denoted as $|G|$. For convenience, we will assume all sub-conferences in a scenario are nonempty set. In other words, a scenario is defined as $S=\{G_i: |G_i|>0 \text{ and } 1 \leq i \leq N\}$. We will name user $i^{th}$ with empty sub-conference $G_i=\varnothing$ as an idle user. The set $U_S = \{i : G_i = \phi \text{ and } 1 \leq i \leq N\}$ is named as the idle set of scenario $S$. Similarly, a user whose video is viewed in a nonempty sub-conference is named as a busy user. Since each busy user is corresponding to one sub-conference in the scenario, the number of busy users equal to $|S|$, which is the number of sub-conferences in the given scenario S. Based on the fact that a user is either busy or idle, we have the following theorem on video conference $C(N,k)$:

Theorem 1: For any given scenario S of $C(N,k)$ conference, we have

$$|U_S| = N - |S| = \sum_{i \in N - U_s} \left( \frac{1}{k} |G_i| - 1 \right) \quad (1)$$

Proof:
Since a user is either an idle user or a busy user, thus $|N|=|S|+|U_S|$. In addition, every user has to view exactly $k$ videos, which mean each user has to be in $k$ different sub-conference. Hence $kN = \sum_{G_i \in S} |G_i|$. So we have $N = \sum_{G_i \in S} \frac{1}{k}|G_i| = |S| + |U_S|$, or

$$|U_S| = N - |S| = \sum_{i \in N - U_s} \frac{1}{k}|G_i| - |S| = \sum_{i \in N - U_s} \left( \frac{1}{k}|G_i| - 1 \right)$$

In the remaining of the paper, we will focus on studying the conference of $C(N,1)$.

Theorem 2: For a given scenario $S$ in $C(N,1)$ conference, there is a partition $P=\{P_i: 1 \leq i \leq N \text{ and } G_i \in S\}$ in the idle user set $U_S$ that satisfies the following conditions:

1). $|P_i| = |G_i| - 1$, for $\forall G_i \in S$

2). $\bigcup_{\{i:G_i \in S\}} P_i = U_S$

3). $P_i \cap P_j = \varnothing, i \neq j$

Furthermore, let $Q_i = G_i \cap U_S$ that is the set of idle users in the sub-conference $G_i$, following requirements can also be satisfied by the partition $P$:

4). $Q_i \subseteq P_i$, for $|Q_i| < |G_i|$

5). $P_i \subset Q_i$, for $|Q_i| = |G_i|$

6). $\bigcup_{\{i:|Q_i|<|G_i|\}} (P_i - Q_i) = \bigcup_{\{i:|Q_i|=|G_i|\}} (Q_i - P_i)$

Before the proof, let's first explain the meaning of this theorem. In a conference, the upload bandwidth of idle users is the deployable resources in helping to distribute video stream in each sub-conference. In the next section, we will use set $P_i$ in partition $P$ constructed in theorem 2 to help delivering the video in sub-conference $G_i$.

Point 1) in the theorem 2 claims that idle users can be "proportionally" assigned to each sub-conference, more precisely, a sub-conference with $|G_i|$ users could find $|G_i|-1$ idle users as helper. Point 2). and 3）shows the characteristic of partition P.

4), 5) and 6) shows a method to construct partition $P$. Specifically, Points 4) and Points 5) shows the initialization of partition $P$. Points 4) in the theorem 2 shows that, all of the idle users in a sub-conference $G_i$ will be assigned to their sub-conference if the number of idle users in this sub-conference is less than $|G_i|$; otherwise, one idle user should be take out from local sub-conference by the point 5) in the theorem. Finally, point 6) shows that point 1) can be satisfied by distributing the users taken out in point 5 to sub-conference whose number of idle users is less than $|G_i|$ in point 4).

Proof: Point 1) comes directly from the theorem 1. Specifically, let k=1 in equation (1), we have $|U_S| = N - |S| = \sum_{i \in N - U_s} (|G_i| - 1)$. That means there is a partition that satisfies Point 1).

2) and 3）are required by the definition of partition. 4) and 5) are self-evident by the initialization of partition $P$.

We now prove 6). Since each user can view only one video in conference $C(N,1)$, the different sets $G_i$ in scenario $S$ disjoint each other, so do the set $Q=\{Q_i=G_i \cap U_S: G_i \in S\}$. In each sub-conference satisfying $|Q_i|=|G_i|$, we will move a user in $Q_i$ to a set named as $W$ and then assign the remained users of $Q_i$ into $P_i$. Thus we have $|P_i|=|G_i|-1$ and $W = \bigcup_{\{i:|Q_i|=|G_i|\}} (Q_i - P_i)$ in this case.

As to the sub-conference in which $|Q_i| < |G_i|$, we assign $Q_i$ to $P_i$ as self-helper for each sub-conferences. Thus the set of all idle users will be written as

$$U_s = \bigcup_{\{i:G_i \in S\}} Q_i = \left( \bigcup_{\{i:|Q_i|<|G_i|\}} Q_i \right) \bigcup \left( \bigcup_{\{i:|Q_i|=|G_i|\}} P_i \right) \bigcup W$$

The number of users in those sets is then
$$|U_s| = \sum_{\{i:|Q_i|<|G_i|\}} |Q_i| + \sum_{\{i:|Q_i|=|G_i|\}} |P_i| + |W| = \sum_{\{i:|Q_i|<|G_i|\}} |Q_i| + \sum_{\{i:|Q_i|=|G_i|\}} (|G_i|-1) + |W|$$

Based on theorem 1 we have $|U_s| = \sum_{\{i:G_i \in S\}} (|G_i|-1)$, thus

$$|U_s| = \sum_{\{i:|Q_i|<|G_i|\}} (|G_i|-1) + \sum_{\{i:|Q_i|=|G_i|\}} (|G_i|-1) = \sum_{\{i:|Q_i|<|G_i|\}} |Q_i| + \sum_{\{i:|Q_i|=|G_i|\}} (|G_i|-1) + |W|$$

After simplification we have
$$|W| = \sum_{\{i:|Q_i|<|G_i|\}} (|G_i|-1-|Q_i|).$$

Thus we can find additional users from $W$ to each sub-conference $i$ in which $|Q_i|<|G_i|$ to satisfying $|P_i|=|G_i|-1$. Thus we have proved the point 6) that

$$\bigcup_{\{i:|Q_i|<|G_i|\}} (P_i - Q_i) = \bigcup_{\{i:|Q_i|=|G_i|\}} (Q_i - P_i)$$

We name the partition $P=\{P_i: G_i \in S\}$ satisfying all above conditions as a Peer-Help Module (PHM). The $P_i$ set in the partition is called the help set of the sub-conference $G_i$. In following section we will address the problem that how to distribution the video $i$ in the sub-conference $G_i$ with the help set $P_i$.

## III. ACHIEVABLE RATE AND CAPACITY

Figure 1 shows the video distribution configuration with PHM in the sub-conference $G_i$. The video stream generated by user $i^{th}$ is distributed to users in the sub-conference $G_i$ with the help of idle users set $P_i$ that is consisted by $Q_i$ and $P_i-Q_i$, which is idle users that belong to $G_i$ and not belong to $G_i$ respect. In fact this configuration is a slightly modification of the peer-assisted file distribution system depicted as the figure 8 in [4].

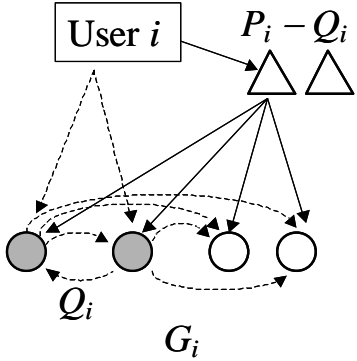

Fig 1 PHM module and peer-assisted file distribution system for the sub-conference $G_i$

Ref.[8] claims that the bottleneck of Internet performance lies in access networks. In addition, Cable networks and DSL networks are two kinds of popular broad access networks today in which the download bandwidth is much larger than the upload bandwidth in general. Hence we will only consider the constraint imposed by the upload bandwidth in the conference system.

Denote $u_i$ and $u(G)$ to be upload bandwidth of user $i^{th}$ and sum of upload bandwidth of the users in a given set $G$ respect. That is

$$u(G) = \sum_{i \in G} u_i$$

Also we define

$$v_i = \frac{u_i + u(P_i)}{|G_i|} - \frac{u(P_i - Q_i)}{|G_i|^2}$$

**Theorem 3.** In $C(N,1)$ conference, the video rate $r_i$ which is $r_i = \min\{u_i, v_i\}$ is achievable for the sub-conference $G_i$ with the $P_i$ in the PHM.

Proof:

Define the threshold $\Delta_i$ as:

$$\Delta_i = \frac{u(Q_i)}{|G_i|-1} + \frac{u(P_i-Q_i)}{|G_i|},$$

In the Appendix we will show that $v_i - u_i = \frac{(|G_i|-1)}{|G_i|}(\Delta_i - u_i)$. Thus $r_i=u_i$ and $r_i=v_i$ is equivalent to $\Delta_i \geq u_i$ and $\Delta_i \leq u_i$ respectively.

In case of $\Delta_i \geq u_i$, user $i$ will divides its video stream into $|P_i|$ substreams and send each substream to one user in $P_i$. Each user in $P_i$ will duplicate the substream it received and retransmit a copy to all other users in the sub-conference $G_i$. User $i$ send a substream to user $j$ in $P_i$ at rate $r_{ij}$ which satisfies

$$r_{ij} = \begin{cases} \dfrac{u_i u_j}{\Delta_i(|G_i|-1)}, & \text{user } j \in Q_i \\ \dfrac{u_i u_j}{\Delta_i |G_i|}, & \text{user } j \in P_i - Q_i \end{cases} \quad (2)$$

Now we check the feasibility of above rate assignment. First, we have $\sum_{j \in P_i} r_{ij} = u_i$ which can be valid by replacing (2) into this expression. Based on this expression, user $i$ is capable of sending out those substreams by using up its uplink's bandwidth. Second, for user $j \in Q_i$, we have $(|G_i|-1)r_{ij} = \frac{u_i u_j}{\Delta_i} \leq u_j$ which can be valid by replacing (2) into this expression and noting the condition $\Delta_i \geq u_i$. Thus all users in $Q_i$ are also capable to duplicate and re-transfer the received substream to other users. Similarly, all users in $P_i-Q_i$ are also capable of transfering the received substream to all other users in the $G_i$. At last, since each user in $G_i$ can receive all substreams generated by user $i$, the video rate of $G_i$ is $r_i=u_i$.

In case of $\Delta_i \leq u_i$, user $i$ will divides its video stream into $|P_i|+1$ substreams. Besides sending a substreams to each user in $P_i$, user $i$ will send an additional substream named as substream 0 to all users in $G_i$. Each user in $P_i$ will also duplicate the substream it received and retransmit them to all other users in the sub-conference $G_i$. The rate $r_{ij}$ is assigned according the following rules, where user j belongs $P_i$

$$r_{ij} = \begin{cases} \dfrac{u_j}{(|G_i|-1)}, & j \in Q_i \\ \dfrac{u_j}{|G_i|}, & j \in P_i - Q_i \end{cases}.$$

And the rate $r_0$ of substream 0 satisfies

$$r_0 = \frac{u_i - \Delta_i}{|G_i|}.$$

Now we check the feasibility of above rate assignment. To begin with, Since $\sum_{j \in P_i} r_{ij} + |G_i| r_0 = u_i$, user $i$ is capable of sending out those substreams. In addition, since $(|G_i|-1)r_{ij} = (|G_i|-1)\frac{u_j}{|G_i|-1} \leq u_j$ for $j \in Q_i$ and $|G_i|r_{ij} = |G_i|\frac{u_j}{|G_i|} \leq u_j$ for $P_i-Q_i$, all users in $P_i$ are also capable of transferring the received substreams to all other users in the $G_i$.

Since each substream is received by each user in $G_i$, the video rate $r_i$ is

$$r_i = \sum_{j \in P_i} r_{ij} + r_0 = \Delta_i + \frac{u_i - \Delta_i}{|G_i|} = \frac{u_i + (|G_i|-1)\Delta_i}{|G_i|} = v_i$$

In the remaining of the paper, we will assume all users in the conference have same upload bandwidth and assume

the upload bandwidth is 1 without loss generality. In this case, we have following corollary

**Corollary:** If $u_i=1$ for each $1\leq i\leq N$, then for a given scenario $S=\{G_i\}$ with the PHM $P$,

1). The video rate
$$r_i = \frac{1+|P_i|}{|G_i|} - \frac{|P_i|-|Q_i|}{|G_i|^2} = 1 - \frac{1}{|G_i|} + \frac{1}{|G_i|^2} + \frac{|Q_i|}{|G_i|^2} \quad (3)$$
is achievable for the sub-conference $G_i$.

2). Furthermore, we have the lower bound
$$r_i \geq 1 - \frac{1}{|G_i|} + \frac{1}{|G_i|^2}, \forall G_i \in S$$

3). $r_i \geq \frac{3}{4}$, for $\forall G_i \in S$

Proof:
1) can be proofed by setting $u=1$ in theorem 3.
2) The minimum $|Q_i|$ is 0 in equation (3), so we have proved the lower bound $1 - \frac{1}{|G_i|} + \frac{1}{|G_i|^2}$.

3) Let $f(x) = 1 - x^{-1} + x^{-2}$, then $f' = x^{-2} - 2x^{-3} = x^{-3}(x-2)$. Since $f(x)$ is a decreasing and increasing function when $x\leq 2$ and $x\geq 2$ respectively. We have $\min f(x) = 1 - 2^{-1} + 2^{-2} = \frac{3}{4}$

From 3), we can reach the following important insight: there is an nontrivial lower bound of the video rate unrelated to specific configuration of sub-conference or number of users in $C(N,1)$. This is very important in the implementation of a conference system. Since a user may join or leave a sub-conference frequently, it is very hard to change the video rate of a sub-conference dynamically according to the number of users in sub-conference. So, we wish to use a fixed video rate in all cases, regardless different sub-conference size $|G_i|$, different number of inside helpers $|Q_i|$, and even different number of users $N$ in the conference. We name the maximum value of such rate as the capacity of conference $C(N,k)$ and this value is denoted as $c_{N,k}$. Specifically,

Definition: $c_{N,k}=\max\{c: c$ is low bound of achievable rate $r_i$ of $G_i$, where $G_i$ is a possible sub-conference of scenarios $S$ in $c_{N,k}\}$.

Based on theorem 3, the achievable rate of a conference is determined by the configuration of sub-conference of $G_i$ and is independent to $N$, which is the number of users in a conference. Thus, it is reasonable to define capacity of conference $C(N,k)$ as: $c_k=\min_N\{c_{N,k},k\}$.

Theorem 4. Under the assumption that all users have an upload bandwidth of 1, the capacity $c_1$ is lower bounded by 5/6:
$$c_1 \geq \frac{5}{6}$$

Proof: In the configuration of Theorem 3, the video owner uses up its upload bandwidth to distribute the video stream to other users in its sub-conference. Here we will use a slightly different video distribution configuration to achieve the lower bound of the capacity $c_1$.

In this configuration, all busy user will use rate $w_i$ to upload the video generated by itself while the remained upload bandwidth $1-w_i$ is used to help distribution the video of other users, which mean a busy user contributes $w_i$ of its upload bandwidth as a video generator and $1-w_i$ of its upload bandwidth as a helper. In other side, an idle user will contribute its full upload bandwidth to help transmitting the video assigned to it. Under this configuration, besides the helpers, the busy users in $G_i$ also contribute upload rate.

Let
$$v_i(w_i) = \frac{w_i+(1-w_i)|G_i-Q_i|+|P_i|}{|G_i|} - \frac{|P_i-Q_i|}{|G_i|^2}$$

With similarly analysis of theorem 3, we can proof the video rate $r_i$ also is $r_i = \min\{w_i, v_i\}$.

If we chose $w_i$ to satisfy the following equation
$$r_i = w_i = v_i(w_i)$$

Thus we have
$$r_i = \frac{r_i+(1-r_i)|G_i-Q_i|+|P_i|}{|G_i|} - \frac{|P_i-Q_i|}{|G_i|^2} = \frac{r_i+(1-r_i)(|G_i|-|Q_i|)+|G_i|-1}{|G_i|} - \frac{|G_i|-1-|Q_i|}{|G_i|^2}$$
$$= \frac{r_i(1-|G_i|+|Q_i|)+2|G_i|-|Q_i|-1}{|G_i|} - \frac{1}{|G_i|} + \frac{1}{|G_i|^2} + \frac{|Q_i|}{|G_i|^2} = \frac{(1-|G_i|+|Q_i|)}{|G_i|}r_i + 2 - \frac{2}{|G_i|} + \frac{1}{|G_i|^2} - \frac{|Q_i|}{|G_i|} + \frac{|Q_i|}{|G_i|^2}$$

Rewrite above equation we have
$$\frac{2|G_i|-1-|Q_i|}{|G_i|}r_i = 2 - \frac{2}{|G_i|} + \frac{1}{|G_i|^2} - \frac{|Q_i|}{|G_i|} + \frac{|Q_i|}{|G_i|^2}$$

After simplification we have
$$r_i = \frac{2|G_i|-2-|Q_i|+\frac{1}{|G_i|}+\frac{|Q_i|}{|G_i|}}{2|G_i|-1-|Q_i|} = 1 - \frac{|G_i|-1-|Q_i|}{(2|G_i|-1-|Q_i|)|G_i|} = 1 - \frac{1}{|G_i|} + \frac{1}{2|G_i|-1-|Q_i|} \geq 1 - \frac{1}{|G_i|} + \frac{1}{2|G_i|-1}$$

Let $f(x) = 1 - x^{-1} + (2x-1)^{-1}$, then
$$f' = \frac{1}{x^2} - \frac{2}{(2x-1)^2} = \frac{(2x-1)^2 - 2x^2}{x^2(2x-1)^2} = \frac{((2+\sqrt{2})x-1)((2-\sqrt{2})x-1)}{x^2(2x-1)^2}.$$

Since $f(x)$ is an increasing function when $x\geq 2$. We have
$$\min f(x) = 1 - \frac{1}{2} + \frac{1}{3} = \frac{5}{6}$$

## IV. CONCLUSION

The achievable rate and the capacity in a MP video conference are addressed and the structure in a conference is studied in this paper. The Peer-Help Module (PHM) in MP video conference is identified and constructed. Peer-assisted file distribution scheme in [4] is applied on the PHM to achieve optimal video delivery in each sub-conference. The capacity of conference is proposed and a lower bound tighter than that implied in [2] is proved. Almost all proofs in this paper are constructive. They can be applied into real implementation directly with a few modifications.

Appendix:

Lemma $\frac{u_i+u(P_i)}{|G_i|} - \frac{u(P_i-Q_i)}{|G_i|^2} - u_i = \frac{(|G_i|-1)}{|G_i|}(\Delta_i - u_i)$

Proof:

$$\frac{u_i+u(P_i)}{|G_i|} - \frac{u(P_i-Q_i)}{|G_i|^2} - u_i = \frac{(|G_i|-1)}{|G_i|}\left(\frac{u(P_i-Q_i)+u(Q_i)}{|G_i|-1} - \frac{u(P_i-Q_i)}{(|G_i|-1)|G_i|} - u_i\right)$$

$$= \frac{(|G_i|-1)}{|G_i|}\left(\frac{u(Q_i)}{|G_i|-1} - \frac{u(P_i-Q_i)}{|G_i|} - u_i\right) = \frac{(|G_i|-1)}{|G_i|}(\Delta_i - u_i)$$